# Probing Electrified Liquid-Solid Interfaces with Scanning Electron Microscopy

*Hongxuan Guo,[1,2,3]† Alexander Yulaev,[2,3]† Evgheni Strelcov,[2,3] Alexander Tselev,[4] Christopher Arble,[2] Andras E. Vladar,[2] John S. Villarrubia,[2] and Andrei Kolmakov[2]**

[1]SEU-FEI Nano-Pico Center, Key Laboratory of MEMS of Ministry of Education, Southeast University, Nanjing 210096, P. R. China

[2]Physical Measurement Laboratory, National Institute of Standards and Technology, Gaithersburg, MD 20899, USA

[3]Maryland Nanocenter, University of Maryland, College Park, MD 20742, USA

[4]Department of Physics and CICECO – Aveiro Institute of Materials, University of Aveiro, 3810-193 Aveiro, Portugal

ABSTRACT

Electrical double layers play a key role in a variety of electrochemical systems. The mean free path of secondary electrons in aqueous solutions is on the order of a nanometer, making them suitable for probing of ultrathin electrical double layers at solid-liquid electrolyte interfaces. Employing graphene as an electron-transparent electrode in a two-electrode electrochemical system, we show that the secondary electron yield of the graphene-liquid interface depends on the ionic strength and concentration of electrolyte and applied bias at the remote counter electrode. These observations have been related to polarization-induced changes in the potential distribution within the electrical double layer and demonstrate the feasibility of using scanning electron microscopy to examine and map electrified liquid-solid interfaces.

INTRODUCTION

The processes related to the formation and dynamics of the electrical double layers (EDLs) at the solid-liquid electrolyte interfaces are crucial to the operation of electrochemical devices,[1] biomedical applications,[2] semiconductor industry processes,[3] electro-kinetic phenomena,[4] (nano-) microfluidics,[5] corrosion, etc. Significant progress has been made in the last two decades both in theoretical understanding and experimental characterization of the structure and properties of the EDLs (see recent review [6] and references therein). A shortage, however, still exists in the experimental analytical tools capable of probing the concentration, composition, and potential profiles within nanometers thin EDLs with sufficient depth and spatial resolution. Of the few existing techniques, scanning probe microscopy (SPM) has demonstrated the ability to interrogate the molecular-level structure of EDL at the interface between the ionic liquid and Au, graphite, or graphene (Gr).[7-9] This method, however, lacks chemical specificity, and molecular dynamics (MD) simulations are usually invoked to facilitate the chemical structure determination. The combination of force curve measurements in atomic force microscopy with infrared[10, 11] or tip-enhanced Raman spectroscopy[12, 13] can be used to resolve this challenge.

The application of traditional surface-sensitive methods such as X-ray photoelectron spectroscopy, atomic emission spectroscopy, near-edge X-ray absorption fine structure spectroscopy *etc.*, that use electrons to probe electroactive liquid interfaces started in the early 80s[14] with the development of differentially pumped electron energy analyzers and has progressed recently through application of synchrotron radiation[15, 16] and novel sample delivery systems.[17-19] These methods take advantage of the nanometer-scale inelastic mean free path of low-energy photoelectrons in liquids,[20] which is on the same length scale as the thickness of the Stern layer in moderately concentrated electrolytes.[21] With the help of these techniques, the variations in composition and electrical potential in EDL of liquid electrolytes has been demonstrated.[21-23] Imaging of such interfaces with low-energy electrons became possible after the introduction of 2D materials as electron-transparent membranes separating the liquid from the vacuum environment.[24-26] In this communication, we are extending the latter approach and demonstrate that standard scanning electron microscopy (SEM) can likewise be used to probe and spatially map changes in the EDL in aqueous electrolytes. The principle of this method is depicted in Figure 1a, b. A low keV electron beam penetrates through the ultrathin bilayer graphene membrane electrode and creates a cascade of electronic excitations and ionization events in the liquid electrolyte close to the electrified interface. Low-energy secondary electrons (SEs) with kinetic energies in the range 0 eV to 50 eV have the escape depths in nanometers in the low single digits[20], so those that escape into the vacuum must originate from the membrane itself or neighboring liquid layers within the EDL (Fig. 1a). Taking advantage of the short escape depth of SEs and the sensitivity of the electron emission yield of the graphene-electrolyte stack to the local potential in the electrolyte (Fig. 1b), we demonstrate the ability of SEM to probe *in situ* the dynamics of the EDL during polarization with high spatial, depth, and temporal resolutions.

RESULTS AND DISCUSSION

Graphene-capped microchannel arrays (MCA) filled with liquid electrolytes have been used in this study.[27] The MCA design considerations for electron spectro-microscopy on liquid electrolytes are reported in [28] and Supplemental Information (SI). To conduct comparative studies of polar and non-polar liquids and electrolytes, we used deionized water (conductivity *ca.* $10^9$ $\Omega\cdot m$), aqueous $10^2$ mol/m$^3$ $CuSO_4$ solution (buffered with 10 mol/m$^3$ $H_2SO_4$), and toluene.

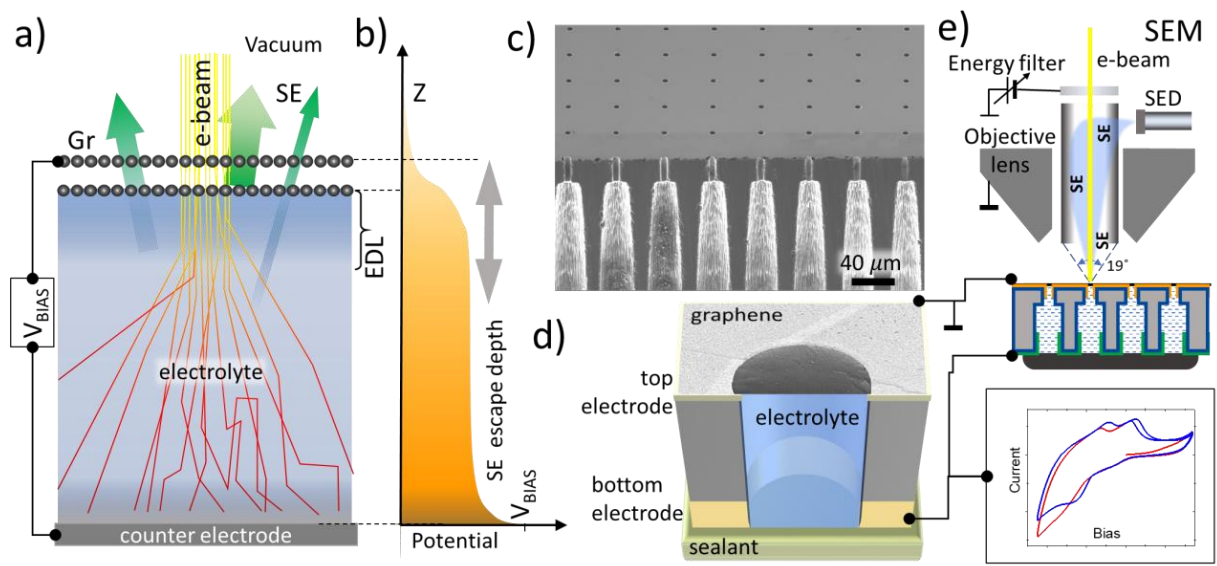

**Figure 1.** A schematic of probing the electrolyte-solid electrified interface with a focused electron beam through a bilayer graphene (Gr) electrode. a) Primary electrons create a cascade of secondary electrons (SE) inside the sample. SEs can escape from the graphene and a thin layer at the graphene-electrolyte interface; b) Biasing the bottom electrode creates an electric double layer (EDL) at the graphene-electrolyte interface which impedes (or promotes) electron emission from the interface. Due to incomplete screening by the graphene, the electrostatic field partially penetrates the vacuum and reduces (or enhances) SE collection efficiency by the electron detector.

c) Cross-sectional SEM image of MCA showing channels inner structure. d) SEM image and wiring diagram of the individual graphene-capped cell filled with electrolyte. e) Experimental setup for SEM imaging during electrochemical cycling. A through-the-lens secondary electron detector (TTL SED) with energy filter is used to discriminate electrons with high kinetic energies and collect only low-energy SE.

A two-electrode MCA electrochemical cell filled with electrolyte (Fig. 1 c, d) was transferred to the SEM chamber (base pressure *ca.* $10^{-4}$ Pa). The cell bias range used for polarization studies was restricted to *ca.* 1 V to avoid water splitting or Cu plating reactions, which manifest themselves as gas bubbles or Cu deposit formation under the graphene. Pt was also selected as a counter electrode to widen the non-Faradaic window. To minimize the effect of the bias on SE trajectories to the detector, the MCA's front Gr reference electrode was grounded, and a polarization bias was applied to the bottom Pt electrode instead (Fig. 1d, e). The magnification, scanning rate, and electron beam current were selected such that several identical channels could be observable in the same field-of-view to improve the data statistics and to avoid radiation damage to the electrolyte. The microscope contrast and brightness settings were kept fixed throughout SEM imaging to ensure reliable comparative analysis. A through-the-lens secondary electron detector with acceptance half angle of *ca.* 19° and an energy filter were employed to collect electrons with energies predominantly between 0 eV and 50 eV (Fig. 1e). Two types of SEM imaging have been used for polarization studies: (i) acquisition of high-resolution individual frames before and after step-like application of the bias across the electrolyte, and (ii) SEM video recording. In both cases, grayscale values of the objects in SEM images, which are proportional to the local secondary electron yield, were analyzed as a function of the bias potential at the bottom electrode and/or time.

To verify the experimental finding, the SE yields were simulated using the JMONSEL (Java Monte Carlo Simulation of Secondary Electrons) simulator (see Methods and SI).[29] We used finite-elements (FE) modeling to perform calculations of distributions of electrical potential and ionic concentration in the microchannels and in the immediate vicinity of the graphene membrane for different electrolyte strengths (see details in SI). In the numerical model, the graphene membrane was modeled as two uncoupled and undoped graphene layers. To calculate the electric screening by the membrane, we used the model developed by Kuroda *et al*.[30]

Figure 2a depicts a typical SEM field-of-view used for these measurements. It contains water-filled MCA channel and one empty reference channel. The bright background with the highest SE yield and corresponding largest signal intensity value $S_{Au}$ is due to the high SE yield of the graphene-covered gold top electrode. The lowest signal $S_G$ originates from graphene which covers an empty (or vapor filled) channel. The channels filled with water can be identified by their intermediate grayscale values due to appreciable (*ca*. 30 %, see discussion below) contribution to the SE yield by the liquid. To make SEM images quantification independent of the instrument/detector specific contrast and brightness settings, we define the effective contrast value $C$ for the graphene covered electrolyte as:

$$C(V) = \frac{S_S(V) - S_G}{S_{Au} - S_G}. \quad (1)$$

Under this definition, the channels' contrast value depends exclusively on the polarization-induced electron yield variations of the electrolyte (and graphene membrane) and is free of the influence of possibly varying image contrast/brightness, beam-induced carbonaceous contamination, *etc*. Figure 2b shows a sequence of SEM images of an individual graphene-covered channel filled with a $10^2$ mol/m$^3$ CuSO$_4$ electrolyte during the application of -0.6 V, 0 V, and +0.6 V bias to the Pt back electrode. The images reveal noticeable and reproducible alternations of the SE signal from the liquid-filled cells upon cell polarization. To quantify the effect of the applied bias on the SEM contrast variation, we plot the difference between the contrast values with and without polarization $\Delta C(V) = C(V) - C(0)$ as a function of bias comparatively for aqueous $10^2$ mol/m$^3$ CuSO$_4$ solution, pure water and toluene (Fig. 2c). The contrast difference, $\Delta C$, decreases quasi-linearly with the electrical bias within the -0.8 V to +0.7 V potential window and becomes non-linear at higher biases presumably due to the onset of Faradaic processes. The temporal SE signal evolution upon the bias onset shows a fast rise followed by saturation (Fig. 2d), which corroborates the redox-free electrolyte polarization process within the selected bias window. Compared to apparent variations of $\Delta C$ for polar liquids, the contrast changes for nonpolar ones (toluene) are negligible (Fig. 2c). Moreover, it was found that toluene-filled cells exhibit a bias-independent SE signal within the widest (-2 V to 2 V) electrochemical window used.

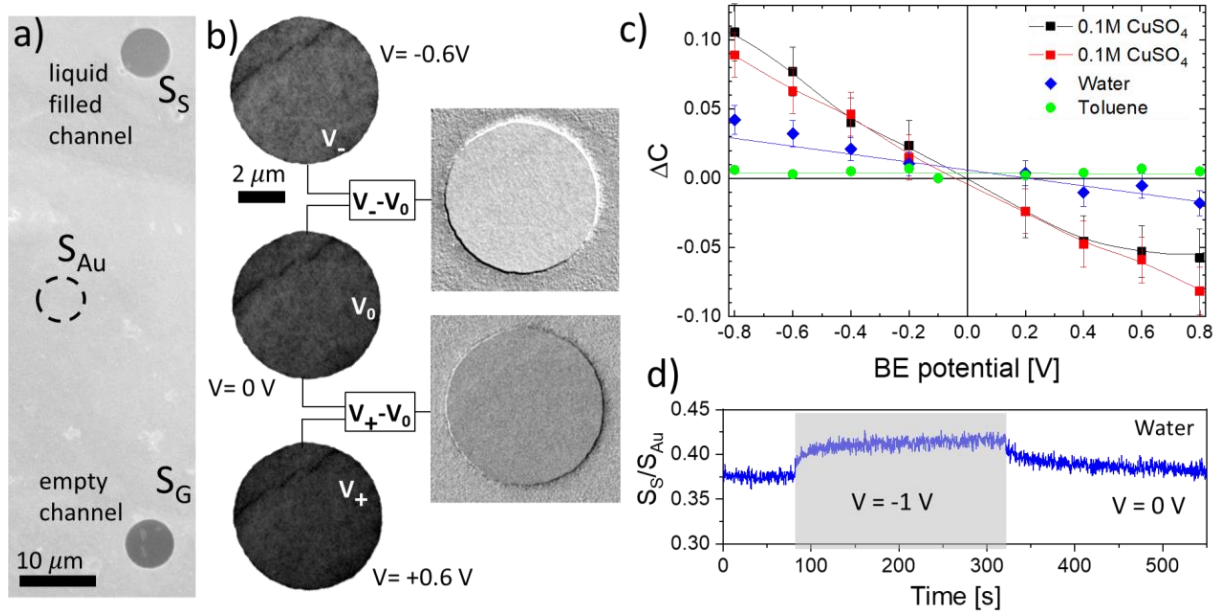

**Figure 2.** Variations of the SE signal acquired from the MCA sample. a) SEM image of an MCA liquid-filled sample with 3 different regions for quantitative analysis of the image contrast: $S_S$ - a liquid sample covered with a Gr membrane, $S_G$ - an empty cell capped with a Gr membrane, and $S_{Au}$ – an area of the Au top electrode covered with Gr. b) SEM images of a cell filled with $10^2$ mol/m$^3$ CuSO$_4$ and capped with a double layer (2L) Gr under -0.6 V, 0 V, and +0.6 V bias applied to Pt counter electrode. All SEM parameters such as contrast/brightness/energy beam current *etc*. were kept fixed throughout image acquisitions. Graphene wrinkles arrear as dark fibrils. Insets are results of the corresponding image subtractions. c) Dependence of the differential SE contrast, $\Delta C$, on an electrical potential on the back electrode for $10^2$ mol/m$^3$ CuSO$_4$ (two different samples shown to demonstrate reproducibility), distilled water, and toluene. Error bars are standard deviations from three measurements on independent channels. d) Temporal evolution of the electrolyte-filled channel contrast upon ON/OFF bias onset.

We attribute the observed dependence of the SE signal of the electrical cell bias to the modulation of the electric potential within the EDL at the graphene-liquid interface, which, in turn, affected the local secondary electron yield. More specifically, we base our conclusions on the following:

An EDL is formed at the graphene-electrolyte interface and the potential profile within the EDL depends on the type and concentration of the electrolyte, applied bias, and geometry of the liquid cell and electrodes (see description below).

Despite the relatively large (*ca*. 200 nm) penetration depth of the primary 2 keV electrons, the SE signal $S_S$ originates from a very narrow, few nanometer thick top layer which includes the bilayer graphene and graphene-electrolyte interface and thus can be defined as $S_S=S_G+S_{EL}$ where $S_{EL}$ is the secondary electron signal from the electrolyte slab. One can exclude the "chemical" component in $S_{EL}$ due to adsorbed ions in our case, since their concentration is about two orders of magnitude lower than that of the solvent, even at a bias of -1 V (see discussion below). Therefore, $S_{EL}$ is determined by the SE yield of the solvent to a large extend.

The electrostatic field induced by electrolyte polarization can be non-zero inside the graphene and cannot be completely screened by bilayer graphene.[30, 31] Based on our prior electron attenuation tests,[26] the realistic bilayer graphene membrane is contaminated and contains additional carbonaceous layer(s), in between and on top of individual graphene sheets and this has been taken into account in the simulations of the SE yield.

The contrast change due to redox reaction like Cu plating is ruled out for the selected bias window since the SE modulations occur much faster (Fig. 2d) than diffusion-controlled Cu deposition and homogeneously over the entire membrane area without the characteristic formation of Cu nucleation centers (see also SI).

To substantiate the above conclusions, we start with FE modeling of the potential distribution inside the electrochemical cell and in the vicinity of the electrified graphene-electrolyte interface as a function of liquid medium type, its concentration, and bottom electrode potential. Figure 3a shows a side-by-side comparison of the calculated distributions of electrical potential for toluene (left half of the channel) and $10^{-3}$ mol/m$^3$ CuSO$_4$ (right half of the otherwise symmetrical channel) under 1 V electrical bias applied to the bottom electrode relative to the grounded top electrode.

With the non-conducting toluene, which does not form an EDL, the potential distribution along the channel is divided into three distinct parts: with the potential of the bottom electrode along the entire length of the inner channel coated with the metal deposit, the intermediate potential in the middle (defined by the floating potential of the Si matrix), and near zero (ground) potential at the very top of the channel (see magnified inset in Fig. 3a) down to the depth of the inner wall metal deposit film of 6 μm. Simulations indicate, therefore, that the graphene and the very top of the liquid at the interface with graphene are effectively at the potential of the top Au electrode, which is zero, independently of the potential on the bottom electrode. This explains the observed independence of the SE contrast with the voltage bias at the bottom electrode for toluene shown in Figure 2c.

In the case of diluted electrolytes and water with impurities (as our case is), the screening by EDLs is incomplete, as seen in the potential map in Figure 3a (right half of the cell) for a concentration of $10^{-3}$ mol/m$^3$, when the Debye length in CuSO$_4$ is about 150 nm. Thus, the "neck" part of the channel is nearly equipotential up to the distance of *ca.* 3 µm below the graphene membrane (see zoomed inset in Fig. 3a) with a potential lower than the potential of the electrolyte bulk. In this range of concentrations, the electrolyte potential at the graphene membrane is closer to the potential of the top electrode. This situation corresponds also to the case of the DI water in the plots in Figure 2c, which is probably contaminated by ions and, therefore, behaves like a very diluted electrolyte.

With increasing ionic concentration, the Debye length of the electrolyte decreases, and the potentials of the silicon substrate and the deposited metal films become increasingly screened by the EDLs. At high salt concentrations, when the EDL thickness becomes on the order of 10 nm or less (for a concentration of $10^2$ mol/m$^3$, the Debye length in CuSO$_4$ is about 0.5 nm), the electrode potentials are completely screened by the Debye layers in the electrolyte. In this case, the electrolyte potential is uniform across the whole electrolyte volume except the EDLs at its boundaries and is determined by the ratio of the EDL layer capacitances at the electrodes, which is equal to the ratio of the areas of the electrode deposited films in contact with the electrolyte. Thus, with the ratio of the electrode areas equal to $\frac{100\ \mu m \times 30\ \mu m}{13\ \mu m \times 6\ \mu m} \approx 40$ (here the top and bottom channel circumferences and metal deposit depth are used), the electrolyte potential is approximately equal to the potential of the larger, bottom, electrode up to the EDL (Fig. 3c). As an example, in Figure 3b, the potential at the electrolyte-graphene interface is 0.6 V on the channel axis, while the electrolyte bulk is at about 0.95 V, which is nearly the potential of the bottom electrode.

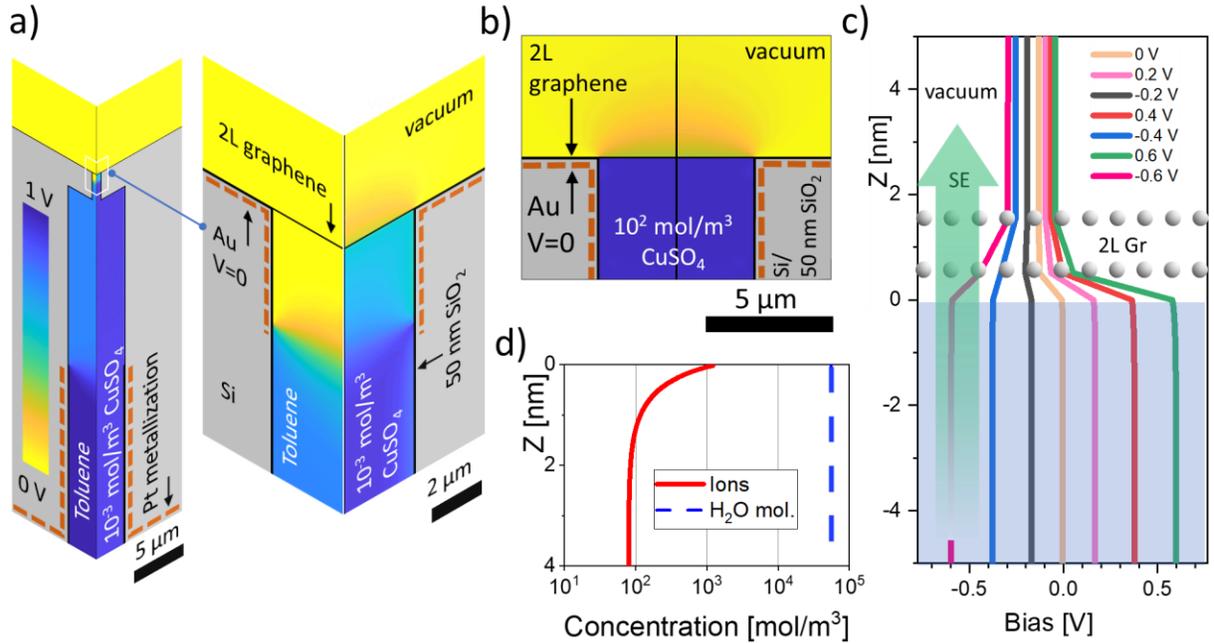

**Figure 3.** FE numerical modeling of the potential distribution in the channels filled with different liquids. a) The comparison of the electrical potential maps inside diluted ($10^{-3}$ mol/m$^3$) CuSO$_4$ electrolyte, and toluene under an electrical bias of 1 V at the bottom electrode. b) The same for a concentrated $10^2$ mol/m$^3$ CuSO$_4$ electrolyte. c) The potential distributions at the graphene -$10^2$ mol/m$^3$ CuSO$_4$ electrolyte interface calculated along the cell central axis and as a function of polarization bias. d) Cation concentration near the two-layer Gr electrode (red curve) compared to the molar concentration of water (blue dashed line) calculated for $10^2$ mol/m$^3$ CuSO$_4$ electrolyte with the bottom electrode biased at 1 V. $z=0$ nm corresponds to the Gr-electrolyte interface.

More precisely, the calculated electric potential distributions along the channel axis in the vicinity of the graphene membrane are shown in Figure 3c. Both the graphene membrane and the liquid layer at the interface have a non-zero potential, which depends on the bias of the bottom electrode. Moreover, due to the limited charge density in the double layer graphene, the electrostatic field of the EDL cannot be screened completely and creates a potential drop between the outer graphene layer and a grounded electron detector. Note that even for concentrated polarized electrolyte, the interfacial concentration of the solute ions is about two orders of magnitude lower than the water molecules concentration (Fig. 3d). Therefore, the observed polarization-dependent SE contrast modulations are due to the potential variations at the graphene and EDL and not chemical composition differences at the interface.

We now discuss the influence of the potential at the liquid-solid interface and in the vacuum on the SE emission signals recorded from the graphene-capped electrolytes. Details of the JMONSEL simulation are provided in the SI. Here we note that bilayer graphene was modeled by an equal thickness graphite slab. Since the measured SEM contrast was defined according to Eq. 1, we normalized JMONSEL simulated SE yields similarly. To do so, we determined for each bias the simulated SE yields for electrons incident on the graphite-covered center of an empty electrochemical cell, on the center of a filled one, and on a gold region 2 µm outside of the edge of a cell (see modeling details in SI). The simulated yield for the $10^2$ mol/m$^3$ CuSO$_4$ cell is shown in Figure 4a where it is compared to the measured values from Figure 2c, both plotted as fractional changes in the yield $\Delta C$ relative to the yield at null bias $C(0)$. The simulation also analyses the effects of finite detector aperture (AP) and unscreened electrostatic field on the SEM contrast changes with the cell bias. Considering the detected SE yield as $\delta = \delta_t p_D$, where $\delta_t$ is the transmitted yield (the fraction of electrons that emerge through the sample surface) and $p_D$ is the probability that an electron, once transmitted, will be detected. The presence and polarity of the electric field at the graphite and graphite-electrolyte interface affects $\delta_t$. In the case of the positive bias, the energy barrier is formed within the EDL (the potential energy difference between an electron inside the electrolyte and one outside, see also Fig. 3c) such that the SE that strike the barrier with insufficient energy are reflected back into the sample (Fig. 4c). With the negative bias, the potential change is negative over a mean free path, which increases the electron's energy and hence the probability of transmission (Fig. 4d). Qualitatively, this mechanism should produce a negative slope in the yield *vs.* bias curve, as observed. Outside the sample, the electrostatic field incompletely screened by graphene affects $p_D$. For example, in the extreme case of a positive surface potential, SE just outside the sample surface have negative potential energy. Those with a kinetic energy insufficient to overcome the potential difference between the surface and grounded detector will not be able to reach the detector. They return and are captured by the sample (Fig. 4c). This mechanism also produces a negative slope, but only for positive surface potential. If the potential is 0 V or less, all electrons escape. However, if the detector is above the sample (as ours is) and has a finite entrance aperture, electrons that would otherwise be emitted over $2\pi$ steradians are focused more tightly in the direction normal to the surface when the bottom electrode potential is negative (Fig. 4d). Thus, with a finite entrance aperture, the external field also is expected to produce an additional negative yield *vs.* bias slope, for negative biases, too.

To evaluate the relative contribution of the aforementioned effects, the SE yield of the graphite-water interface was simulated under different conditions as indicated in the legend of Figure 4a. The curves labeled with the "AP" suffix counted only electrons that hit the detector within a finite entrance aperture (19° half angle). The one without this suffix includes all SEs emitted within $2\pi$ steradians. The curve labeled with "$\vec{E} = 0$" was simulated with the graphite/vacuum interface above the cell artificially shifted to 0 V potential. (The potentials inside the cell were shifted by the same amount, thereby keeping all potential differences, *i.e.*, the electric fields, the same inside the cell.) Thus, all 3 simulated curves have the same electric fields inside the sample, and all have negative slopes. The solid red simulated curve (marked *AP*) includes the bias-induced electric field outside the sample and the finite detector aperture. It is the one that most closely agrees with the experimental data (EXP). The remaining two simulated curves neglect either the external field or the finite aperture contributions. They agree well with each other but differ more strongly from the measurement, especially at higher biases.

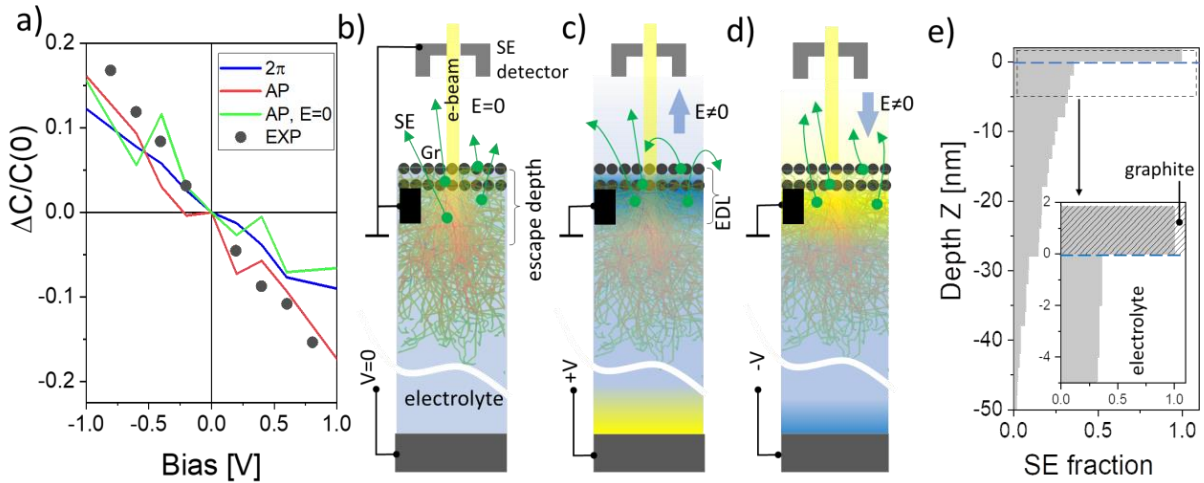

**Figure 4.** JMONSEL simulations of the SE yield. a) Simulated differential SE contrast, *ΔC/C(0)*, as a function of the bottom electrode bias for $10^2$ mol/m$^3$ CuSO$_4$ for the following conditions: unrestricted (2π radians); SE collection (blue curve); electrons collected with 19° angular aperture AP, (red curve); electrons collected with 19° angular aperture under the conditions of complete electrostatic screening of the EDL potential ($\vec{E} = 0$, green curve). Black dots show the experimental data for comparison. b-d) Diagrams showing the effect of the unscreened electrostatic field on the electron collection efficiency (see details in the text). e) The fraction of the SE with the depth of origin deeper than Monte Carlo simulated *via* back-tracing 248 detected SEs (see details in the text).

Finally, we want to comment on the partitioning of the detected SE between the capping electrode and electrolyte and the overall escape depth of the collected SEs. For that, for every simulated detected SE, we determined the z coordinate corresponding to its deepest penetration into the sample. For this purpose, the part of its parent electron's trajectory prior to its own generation is treated as though part of its own trajectory and likewise for earlier ancestors. Electric fields below this z of deepest penetration can have had no influence on this electron's fate. Fields between deepest penetration and the detector have influence but not always much. In particular, electrons at deep parts of this causal chain are likely to have been more energetic and therefore less sensitive to electric fields. For this reason, this metric should be treated as a bound. Figure 4e depicts the fraction (as a function of $z$) of escaped SEs with origin deeper than $z$. Here, zero line is at the capping electrode-electrolyte interface and two-layer graphene (including contaminations) is represented as graphite with a thickness of 1.84 nm above it (see inset in the Fig. 4e). The modeling indicates that about 70 % of the detected SEs originate inside the graphene capping layer, and the remaining 30 % of electrons have "ancestors" distributed along the long tail that extends as deep as 60 nm below the graphene/electrolyte interface. The calculated mean escape depth of such "electrolyte" SE is *ca.* 6.9 nm. Note that the electric field within the graphene is sensitive to the potential variation in the EDL just below it, so even SEs that originate in the graphene layer are sensitive to the applied bias. The method, therefore, is suitable for interrogation of the electrochemical systems with lateral potential gradients and lateral electrode structure inhomogeneities.

CONCLUSIONS

To summarize: we have used liquid electrochemical cells with an electron transparent two-layer graphene electrode to elucidate the effect of the electrolyte polarization at the graphene-liquid interface on the SE emission using SEM. We have found that the SE emission from such an interface is sensitive to the formation and re-structuring of the EDL and can be recorded using typical in-lens SE detector of the SEM. Finite element analysis modeling of the potential distribution near the interface coupled with Monte Carlo simulations explains the general trends of bias-dependent contrast and determines its values. The proposed method is most useful for studying electrochemical systems where laterally heterogeneous interfaces are present and high-resolution polarization mapping is required. As we have shown recently, care has to be taken in these cases to minimize the radiation damage of the electrolyte[32] and damage-free quantification and mapping of the polarization potentials of radiation sensitive electrolytes can be also done using Kelvin Probe Force Microscopy.[33]

METHODS

**Sample preparation**

Fabrication of MCA include the following steps (see detailed procedure in Supporting Information): (i) Bottle-shaped microchannels were formed using reactive ion etching of lithographically defined Si wafer. (ii) 100 nm thick thermal oxide was grown on the Si surface. (iii) 200 nm Au on 10 nm Cr and 30 nm Pt electrodes were deposited onto the MCA's front and back surfaces. (iv) Bilayer Gr was transferred using a PMMA sacrificial layer onto the Au electrode. (v) PMMA was dissolved in hot acetone. (vi) MCA was filled with a liquid electrolyte using a substitutional filling. (vii) Backside openings of the filled channels were sealed using a water immiscible epoxy cured by UV irradiation.

**Monte Carlo simulations**

To determine the SE yield, we simulated 1 million beam electrons incident on each relevant position (*e.g.*, the center of the cell or the graphite-covered gold near the cell). Each electron's trajectory consisted of a series of steps, each of which was undisturbed flight (corrected for the effect of the electric field determined by the finite-element model below) terminated by an "event."

The simulated events included elastic scattering (*i.e.*, Mott scattering using mean free paths from [34]), SE generation using dielectric function theory, longitudinal optical phonon scattering, and refraction or reflection at material interfaces. SEs were tracked in the same way. Tracking ended when an electron's energy fell below the amount required for it to escape the sample, or when it was detected at a distance 0.1 m from the sample. Each detected electron's position, the direction of motion, and energy at detection were logged for subsequent analysis. Details of these models can be found in an earlier paper.[29]

ASSOCIATED CONTENT

**Supporting information**

The Supporting Information is available free of charge on the ACS Publications website at DOI: ---. This includes details on sample preparation, JMONSEL simulator, and finite element modeling of potential distribution inside the channel filled with liquid electrolyte.

AUTHOR INFORMATION

**Corresponding Author**

*E-mail: andrei.kolmakov@nist.gov  ORCID 0000-0001-5299-4121

[†]These authors contributed to the work equally.

**Author Contributions**

A.K. conceived the idea and directed the work. H.G. and A.Y. conducted the SEM polarization experiments and preliminary data analysis. E.S. and C.A. carried out electrochemical measurements. J.V. and A.T. performed Monte Carlo and finite element simulations, correspondingly. A.V. contributed to the data interpretation. The manuscript was compiled by A.K. using contributions of all authors. All authors have given approval to the final version of the manuscript.

ACKNOWLEDGMENT

H.G., A.Y., and E.S. acknowledge support under the Cooperative Research Agreement between the University of Maryland and the National Institute of Standards and Technology Center for Nanoscale Science and Technology, Award 70NANB14H209, through the University of Maryland. In part (A.T.), this work was developed within the scope of the project CICECO-Aveiro Institute of Materials, UIDB/50011/2020 and UIDP/50011/2020, financed by national funds through the Portuguese Foundation for Science and Technology/MCTES. MCA chips have been fabricated by Dr. Brian Hoskins (PML, NIST). High quality graphene was kindly provided by Ivan Vlassiouk (ORNL).

ABBREVIATIONS

SEM, scanning electron microscope; SE, secondary electron; Gr, graphene; 2L, double-layer; TTL detector, through-the-lens detector; MD, molecular dynamics; EDL, electrical double layer; MCA, micro channel array; JMONSEL, Java Monte Carlo Simulation of Secondary Electrons; AFM, atomic force microscopy; SPM, scanning probe microscopy.

SYNOPSIS

The nanometer range escape depth of secondary electrons in aqueous solutions makes SEM a suitable tool for probing ultrathin electrical double layers at the solid-liquid electrolyte interfaces with high spatial and temporal resolutions.

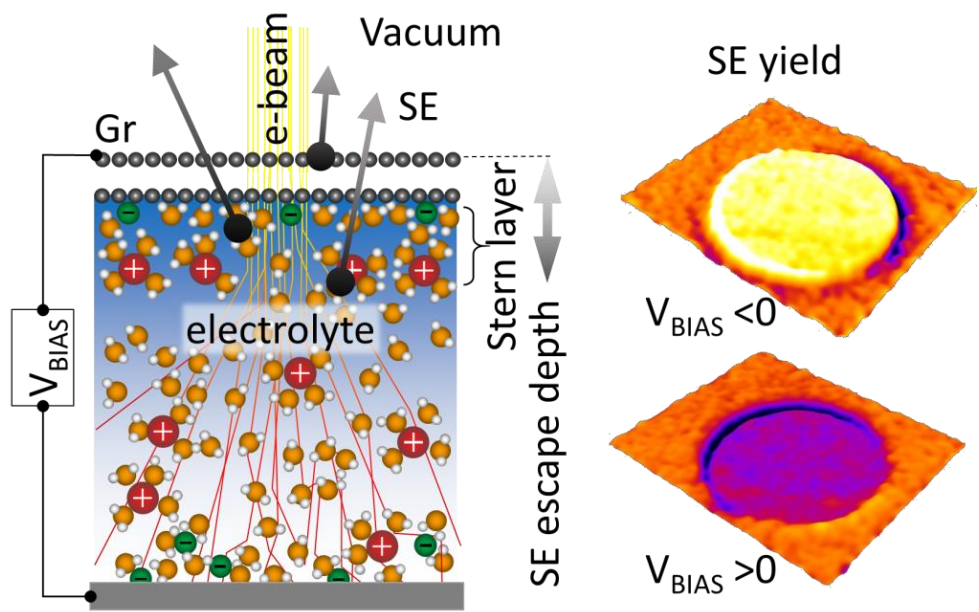

TOC figure